\documentclass[sigconf,nonacm,screen]{acmart}
\usepackage{amsfonts}
\usepackage{amsmath}
\usepackage{amsthm}
\usepackage{array}
\usepackage{bm}
\usepackage{booktabs}
\usepackage{bxtexlogo}
\usepackage{caption}
\usepackage{enumitem}
\usepackage{float}
\usepackage{geometry}
\usepackage{graphicx}
\graphicspath{{figs/}}
\usepackage[utf8]{inputenc}
\usepackage{makecell}
\usepackage{mathtools}
\usepackage{multirow}
\usepackage{natbib}
\setcitestyle{authoryear,numbers,aysep={}}
\usepackage{soul}
\usepackage{stfloats}
\usepackage{subcaption}
\usepackage[switch]{lineno}
\def\innerprod<#1>{\langle #1 \rangle}

\theoremstyle{definition}

\theoremstyle{remark}

\AtBeginDocument{%
  }

\newcommand{\Prob}{\mathbb{P}}
\DeclareMathOperator{\CVaR}{CVaR}

\begin{document}

\title{Financial Tail Risk Beyond Lipschitz Continuity via Semi-Discrete Optimal Transport}

\author{Ryan M. Engel}
\authornote{Joint affiliation with Stony Brook University, USA.}
\affiliation{%
  \institution{Simudyne}
  \country{United Kingdom}}
\email{ryan.engel@simudyne.com}

\author{Kibaek Lee}
\affiliation{%
  \institution{Simudyne}
  \country{United Kingdom}}

\author{Namid Stillman}
\affiliation{%
  \institution{Simudyne}
  \country{United Kingdom}}
\email{namid@simudyne.com}

\renewcommand{\shortauthors}{}

\begin{abstract}

Financial returns are heavy-tailed, and accurate tail risk estimation is central to portfolio risk management. Modern neural generators sample by pushing a simple base distribution through a learned map, and for training stability that map is built from Lipschitz components. This is the binding constraint: a Lipschitz map of a Gaussian is sub-Gaussian, so heavier-tailed targets admit no exact match at any finite Lipschitz constant. The Monge--Amp\`ere equation ties the Brenier map's local distortion to the density ratio $f/(g\circ T)$, so a deeper trough in the target density requires a higher-gain map and yields a higher-variance estimator. The argument needs only bounded distortion, so it covers normalizing flows, flow matching, GANs, and diffusion samplers alike.

Semi-Discrete Optimal Transport (SDOT) relaxes the map's regularity rather than the source's tail class. Its power diagram gives every training observation a cell holding exactly $1/N$ of the source measure, and tail observations are reached by crossing a cell boundary rather than by stretching. Our primary experiment sweeps severity over a calibrated Merton jump-diffusion spanning kurtosis 94 to 1,679. SDOT holds tail ratios at $0.85$--$0.94$ with cross-seed standard deviations below $0.025$, while every learned generator either compresses the tails or inflates them with a variance that grows alongside. Further experiments carry the result to real S\&P~500 returns and to a 21-year backtest, where SDOT gives the best risk-adjusted market-neutral strategy under CVaR optimization (Sharpe $0.70$, max drawdown $-2.60\%$, against $0.40$ for the next-best generator).

\end{abstract}




\maketitle

\section{Introduction}
\label{sec:intro}
Portfolio risk management relies on generating plausible future return scenarios to estimate tail risk measures such as Conditional Value-at-Risk (CVaR), and the accuracy of these estimates depends on the generative model's ability to reproduce the heavy-tailed distributions observed in financial returns. Daily S\&P~500 constituent returns exhibit kurtosis of approximately 88 after standard data cleaning \cite{cont2001empirical}, orders of magnitude above the Gaussian benchmark.

Recent advances in deep generative modeling have produced powerful neural density estimators and samplers \cite{dinh2017density, chen2018neural, lipman2023flow, ho2020denoising, onken2021otflow}. These methods differ in what they require of the transport map $T\!:\mathbb{R}^d\to\mathbb{R}^d$ pushing a simple base distribution to the target: normalizing flows and flow matching constrain $T$ to be a diffeomorphism, enabling exact likelihoods, while adversarial and diffusion samplers carry no such constraint. What they share is that $T$ is built from affine layers and Lipschitz nonlinearities, and is therefore itself Lipschitz. Section~\ref{sec:lipschitz} argues that this shared property, not smoothness or invertibility, is what binds on heavy-tailed targets: a Lipschitz map of a Gaussian is sub-Gaussian, so heavier-tailed targets admit no exact match at any finite Lipschitz constant \cite{jaini2020tails}. What varies is severity, and the Jacobian equation $\det(DT(x)) = f(x)/(g\circ T(x))$ supplies the measure: a deep trough in the target density demands a high-gain map, which yields a high-variance estimator.

Prior work responds by making the source heavier-tailed \cite{jaini2020tails,huster2021pareto}; we relax the regularity of the map instead. Semi-Discrete Optimal Transport (SDOT), grounded in the geometric variational formulation of Gu et al.~\cite{gu2013variational}, solves for a power diagram assigning one convex cell to every training observation, each receiving exactly $1/N$ of the source measure at the optimum. The map is discontinuous across cell boundaries and so escapes the bound above: a tail observation is reached by crossing a boundary rather than by stretching, with its empirical frequency inherited from the data rather than learned against a regularity constraint.

Our contributions are as follows.
\begin{enumerate}[nosep]
    \item Neural generators fail on heavy tails because they are Lipschitz, not because they are smooth or invertible, so the constraint binds equally on flows, flow matching, adversarial, and diffusion samplers.
    \item A quantitative form of the obstruction: prior work shows a Lipschitz map preserves its source's tail class, and we connect the required distortion to the Monge--Amp\`ere density ratio, computing it in closed form to replace a binary reachability statement with a severity that can be swept.
    \item An empirical test of that severity. Across four regimes spanning kurtosis 94 to 1,679, every learned generator degrades in tail ratio, cross-seed variance, or both, while SDOT holds tail ratios at $0.85$--$0.94$.
    \item A 21-year out-of-sample backtest. Bootstrap reproduces the training measure but yields a degenerate zero-weight portfolio under CVaR-minimizing optimization, while SDOT's absolutely continuous scenarios support the best risk-adjusted market-neutral strategy tested.
\end{enumerate}

\section{Related Work}
\label{sec:related}
Normalizing flows~\cite{dinh2017density,kingma2018glow} compose elementary bijections to learn an invertible map between the data and a simple base, giving exact log-likelihoods via the change-of-variables formula. Continuous normalizing flows~\cite{chen2018neural,grathwohl2019ffjord} relax the architectural restrictions of discrete flows by parameterizing a free-form ODE, and flow matching~\cite{lipman2023flow} avoids simulation during training by regressing a velocity field against conditional paths. 

That Lipschitz maps cannot manufacture heavy tails is known in this literature. Jaini et al.~\cite{jaini2020tails} show that a Lipschitz map returns a target with the tail class of its source; their remedy is to learn the source alongside the map, as do Huster et al.~\cite{huster2021pareto} and Laszkiewicz et al.~\cite{laszkiewicz2022marginal} for adversarial and marginal-flow models. Section~\ref{sec:lipschitz} records the statement in a generality covering every generator benchmarked here, quantifies the required constant through the Monge--Amp\`ere density ratio, and takes the complementary route of relaxing the map rather than the source.

Denoising diffusion probabilistic models~\cite{ho2020denoising,song2021score} instead learn to reverse a fixed Gaussian noising process, generating by denoising step-by-step from noise. The reverse process is stochastic rather than diffeomorphic, but each step is Lipschitz, so the sampler is a Lipschitz function of the injected noise and Section~\ref{sec:lipschitz} applies. Generative adversarial networks~\cite{goodfellow2014gan} have been applied to asset return and volatility simulation~\cite{wiese2020quantgan}. Tail-GAN~\cite{cont2023tailgan} trains against a joint scoring function for VaR and expected shortfall, so that simulated paths reproduce the tail risk of a prescribed set of benchmark strategies. Both modify the training objective without changing the geometry of the map, which remains a Lipschitz pushforward of a simple base. An et al.~\cite{an2020aeotgan} identify the same discontinuity as the source of mode collapse, through the lens of Figalli's regularity theory for optimal maps between non-convex domains (Section~\ref{sec:regularity}).

Classical optimal transport~\cite{Villani2009} seeks the minimum-cost map between two distributions, with computational approaches including entropic and discrete solvers~\cite{cuturi2013sinkhorn,flamary2021pot}, neural approximations of the Monge map~\cite{makkuva2020optimal}, and sliced formulations~\cite{bonneel2015sliced}. OT-Flow~\cite{onken2021otflow} applies these ideas to flow-based generators by penalizing a CNF velocity field to encourage straight trajectories. Semi-discrete OT~\cite{gu2013variational,merigot2011multiscale} partitions the source domain into power cells, each assigned to a discrete target point, and AE-OT~\cite{an2020aeot} scales the formulation to high dimensions via Monte Carlo volume estimation. Imposing no continuity constraint, the power diagram represents discontinuity explicitly through its cell boundaries, with coverage following from the optimality condition rather than from training, as Engel et al.~\cite{engel2025geometric} demonstrate across several datasets.

\section{Background}
\label{sec:background}
This section covers the theoretical foundation for the proposed methodology, including the theorems for optimal transportation, the Monge-Amp\`ere equation, and the variational approach to solve the semi-discrete optimal transportation problem.

\subsection{Optimal Transportation}
\label{sec:ot}
The prescribed Gauss curvature problem, the local form of the classical Minkowski problem \cite{Minkowski1897}, seeks a convex function $u(x)$ on $\Omega$ with given curvature $K(x)$, governed by the Monge--Amp\`ere equation
\begin{equation}
    \frac{\det(D^2u)}{(1+|\nabla u|^2)^2} = K(x).
    \label{eq:prescribed_gauss_curvature}
\end{equation}
Its Legendre dual
\begin{equation}
    u^*(y) = \sup_{x\in \Omega} \langle x, y \rangle - u(x)
    \label{eq:legendre_dual}
\end{equation}
satisfies $y = \nabla u(x)$ and $(D^2u(x))^{-1} = D^2u^*(y)$, so the two graphs have reciprocal curvatures. This duality underlies the variational formulation of Section~\ref{sec:geometric}.

The Minkowski problem and the Monge-Amp\`ere equation are closely related to the optimal transportation problem, which seeks the most economical way to transport one probability measure to the other. Suppose the source distribution is $\mu$ with a convex support $\Omega\subset\mathbb{R}^n$, the target distribution is $\nu$ with support $\Omega^*\subset\mathbb{R}^n$. A map $T:\Omega\rightarrow \Omega^*$ is measure-preserving, denoted as $T_\#\mu = \nu$, if for any Borel set $B\subset \Omega^*$, $\int_{T^{-1}(B)} d\mu(x) = \int_B d\nu(y)$.

The cost function $c:\Omega\times\Omega^*\to \mathbb{R}$ measures the cost for moving a unit mass from $x\in \Omega$ to $T(x)\in \Omega^*$. The optimal transportation problem, originally posed by Gaspard Monge in 1781, seeks the measure-preserving map with the minimal total transportation cost: $\min_{T_\#\mu = \nu} \int_\Omega c(x,T(x))d\mu(x)$. Kantorovich relaxed the transportation map to a joint distribution $\rho$ with marginals $\mu$ and $\nu$ gives the Kantorovich problem, whose dual is
\begin{equation}
    \max_{\varphi} \int_\Omega \varphi(x)d\mu(x) + \int_{\Omega^*} \varphi^c(y) d\nu(y),
    \label{eq:kantorovich-dual}
\end{equation}
where $\varphi^c:\Omega^*\to\mathbb{R}$ is the c-transform of $\varphi:\Omega\to\mathbb{R}$, $\varphi^c(y) := \inf_{x\in \Omega} c(x,y) - \varphi(x)$. When the cost is the squared Euclidean distance $c(x,y)=\frac{1}{2}|x-y|^2$, and the source density function is absolutely continuous, then Brenier proved in \cite{brenier1991polar} that the optimal transport map $T$ is the gradient map of a convex function $u:\Omega\to \mathbb{R}$, $T=\nabla u$, $u$ satisfies the Monge-Amp\`ere equation:
\begin{equation}
    \text{det}(D^2u) (x) = \frac{f(x)}{g\circ\nabla u(x)},
    \label{eq:monge-ampere}
\end{equation}
with natural boundary condition $\nabla u(\Omega) = \Omega^*$, where $f,g$ are the density functions of $\mu$ and $\nu$. Comparing Eqn.~\ref{eq:prescribed_gauss_curvature} and Eqn.~\ref{eq:monge-ampere} reveals the intrinsic relation between the Minkowski problem and optimal transportation.

\subsection{Geometric Variational Method}
\label{sec:geometric}

\begin{figure}[t]
    \centering
    \includegraphics[width=0.95\linewidth]{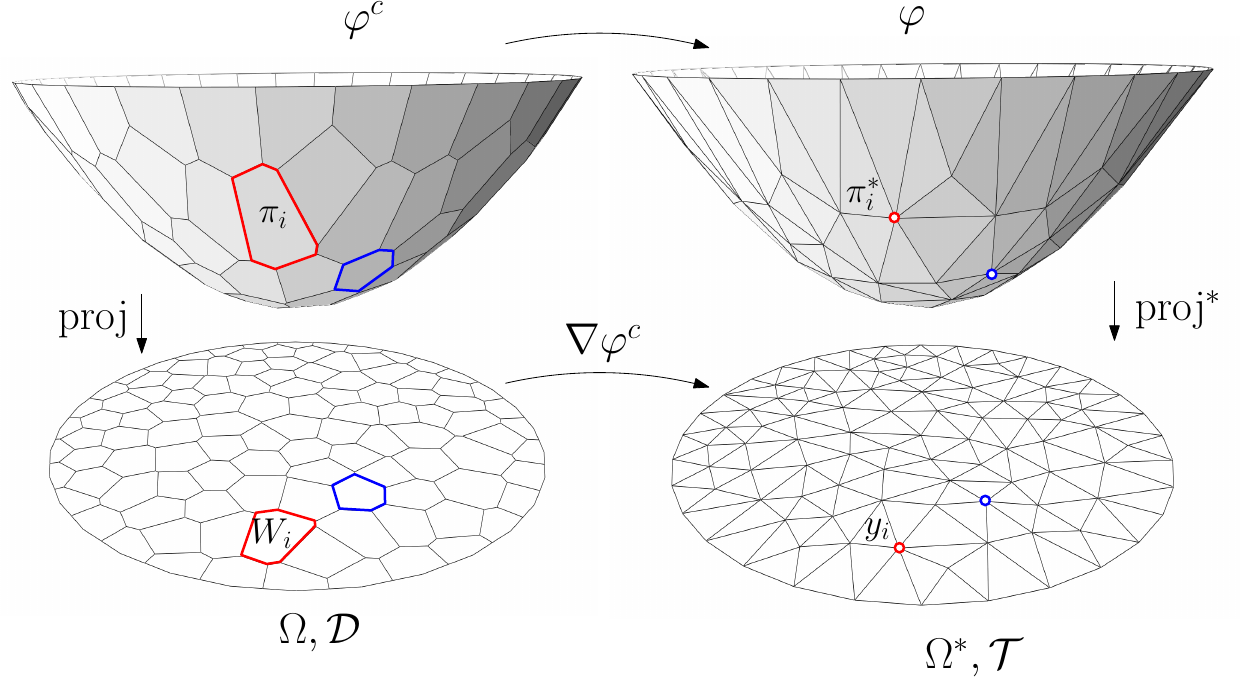}
    \caption{Variational method for optimal transport.}
    \label{fig:PL_duality}
\end{figure}

Gu et al.\ developed a geometric variational method to solve the semi-discrete optimal transport problem, which directly solves the prescribed Gauss curvature problem~\eqref{eq:prescribed_gauss_curvature}. For the squared Euclidean cost $c(x,y)=\frac{1}{2}|x-y|^2$ and under the condition $T_{\#}\mu = \nu$, the quantity $\int_\Omega |T(x)|^2 d\mu(x) = \int_{\Omega^*} |y|^2 d\nu(y)$ is independent of the transport map, so Monge's problem can be reformulated with the cost $c(x,y)=\langle x,y \rangle$ as $\min_{T_\#\mu=\nu}\int_\Omega \frac{1}{2}|x-T(x)|^2 d\mu(x) \iff \max_{T_\#\mu=\nu} \int_\Omega \langle x, T(x)\rangle d\mu(x)$, under which the Kantorovich dual problem~\eqref{eq:kantorovich-dual} becomes
\begin{equation}
    \min_\varphi \int_\Omega \varphi^c(x)d\mu(x) + \int_{\Omega^*}\varphi(y) d\nu(y),
    \label{eq:kantorovich-energy}
\end{equation}
and the $c$-transform reduces to the conventional Legendre dual~\eqref{eq:legendre_dual}.

In practice the target measure is approximated as a sum of Dirac measures, as shown in Fig.~\ref{fig:PL_duality}, $\nu = \sum_{i=1}^n \nu_i \delta(y-y_i),\quad \Omega^*=\{y_1,y_2,\dots,y_n\}$, with Kantorovich potential $\varphi(y) = \sum_{i=1}^n h_i \delta(y-y_i)$. Writing $h=(h_1,h_2,\dots,h_n)$, the graph of $\varphi$ is the convex hull $\text{Conv}(h)$ of the points $\{(y_1,h_1),\dots,(y_n,h_n)\}$, and its Legendre dual $\varphi^c(x) = \sup_{y\in\Omega^*} \langle x,y\rangle - \varphi(y) = \max_{i=1}^n \{\langle x,y_i\rangle - h_i \}$, has the upper envelope $\text{Env}(h)$ of the planes $\pi_i(x)=\langle x,y_i \rangle - h_i$. The projection of the upper envelope induces a power diagram $\mathcal{D}(h)$ on $\Omega$, $\Omega = \bigcup_{i=1}^n W_i(h)$, where $W_i(h) = \left\{ x\in \Omega: \langle x, y_i \rangle - h_i \ge \langle x, y_j \rangle - h_j, \forall j \right\}$. Its dual is the power Delaunay triangulation $\mathcal{T}(h)$ of $\Omega^*$, in which each cell $W_i(h)$ corresponds to the point $y_i$ and each edge $e_{ij} = W_i(h)\cap W_j(h)$ corresponds to the edge connecting $y_i$ and $y_j$. This dual structure supplies the neighbouring cells used for generation in Section~\ref{sec:method}.

The piecewise linearity of $\varphi^c$ gives the dual energy~\eqref{eq:kantorovich-energy} a simple form,
\begin{equation}
E(h) = \sum_{i=1}^n \int_{W_i(h)} (\langle x,y_i \rangle - h_i) d\mu(x) + \sum_{j=1}^n h_j\nu_j,
\label{eq:energy}
\end{equation}
with gradient $\nabla E(h) = (\nu_1 - w_1(h), \dots, \nu_n - w_n(h))$, where
\begin{equation}
w_i(h) = \mu(W_i(h)) = \int_{W_i(h)} f(x) dx,
\label{eq:gradient}
\end{equation}
is the $\mu$-measure of the power cell $W_i(h)$. On the admissible space $\mathcal{H}=\left\{h\in \mathbb{R}^n: \textstyle\sum_{i=1}^n h_i = 0\right\} \cap \left\{ W_i(h) \neq \emptyset, \forall i \right\}$,
the Hessian of $E$ is diagonally dominant with a one dimensional null space $\lambda(1,1,\dots,1)^T$, so $E$ is strictly convex on $\mathcal{H}$ and its minimizer $h^*$ is unique and attained in the interior \cite{gu2013variational}. At $h^*$ the gradient vanishes, so the $\mu$-volume of every cell equals its prescribed measure,
\begin{equation}
    w_i(h^*) = \nu_i = \frac{1}{n} \quad \text{for all } i = 1, \ldots, n.
    \label{eq:optimality}
\end{equation}
The optimal transport map $T$ sends each cell $W_i(h^*)$ to $y_i$. Geometrically, when $\mu$ is the Lebesgue measure, $\text{Conv}(h^*)$ solves the prescribed Gauss curvature problem and hence the Monge--Amp\`ere equation~\eqref{eq:monge-ampere}.

\subsection{Regularity of Transport Maps}
\label{sec:regularity}
The regularity theory of optimal transport is closely connected to the regularity theory of the Monge--Amp\`ere equation. The analysis of this equation builds on the weak solution theory introduced by Aleksandrov through the Monge--Amp\`ere measure \cite{Alexandrov1958}. Caffarelli established fundamental interior regularity results showing that if the densities are bounded away from zero and infinity then convex solutions are locally $C^{1,\alpha}$, and if the densities are H\"older continuous then the potential is locally $C^{2,\alpha}$ \cite{Caffarelli1990,Caffarelli1992}. For the global problem, if the domains $\Omega$ and $\Omega^*$ are bounded, uniformly convex, and have $C^{2}$ boundaries, and if the densities satisfy $0<\lambda\le\rho,\sigma\le\Lambda$ with $\rho,\sigma\in C^{\alpha}$, then Caffarelli proved boundary regularity and obtained $u\in C^{2,\alpha}(\overline{\Omega})$, implying that the optimal transport map is a $C^{1,\alpha}$ diffeomorphism between the domains \cite{Caffarelli1996}.

For general cost functions the MTW condition governs regularity \cite{MaTrudingerWang2005,Loeper2009}, and comprehensive treatments appear in Villani and Figalli \cite{Villani2009,Figalli2017}. Figalli showed that when the target domain is not convex, global regularity may fail. Even for smooth positive densities, singularities of the transport map can appear, although the map remains smooth outside a lower-dimensional singular set \cite{figalli2010regularity}.

Caffarelli's global boundary-regularity theorem \cite{Caffarelli1996} rests on two hypotheses, uniform convexity of the domains, and densities bounded away from zero and infinity. Failure of the first is topological, and yields the coverage argument of Engel et al.~\cite{engel2025geometric}, resting on Figalli's singularity result above, at finite sample size the empirical target is atomic, and a generator with a connected latent space reaches its modes only by traversing the gaps between them.

Failure of the second is metric, and is the mechanism this paper takes as central. The calibrated jump-diffusion of Section~\ref{sec:phase2} has full support and a convex domain, so the topological obstruction does not arise there; what its density violates is the lower bound, and the consequence is not that a continuous map fails to exist but that any such map must be violently distorted. Section~\ref{sec:lipschitz} makes this precise. The semi-discrete formulation is immune to both, since the power diagram represents discontinuities explicitly and imposes no bound on distortion.

\subsection{Lipschitz Constraints on Learned Maps}
\label{sec:lipschitz}
A network of affine layers and Lipschitz nonlinearities is globally Lipschitz, with $\mathrm{Lip}(G)\le\prod_i\|W_i\|_2$, and the devices that stabilize training act to keep this constant small. Bounded distortion, rather than smoothness or invertibility, is what limits the tails such a map can produce.

The constraint follows from Gaussian concentration. If $Z\sim\mathcal{N}(0,I_k)$, $G$ is $L$-Lipschitz and $\psi:\mathbb{R}^d\to\mathbb{R}$ is $1$-Lipschitz, then $\psi\circ G$ is $L$-Lipschitz and the Borell--TIS inequality \cite{borell1975brunn,ledoux2001concentration} gives
\begin{equation}
    \Prob\big(|\psi(G(Z))-m|>t\big)\;\le\;2\exp\!\left(-t^{2}/2L^{2}\right)
    \label{eq:subgaussian}
\end{equation}
for $m$ a median of $\psi(G(Z))$. Every marginal of $G_\#\mu$ is therefore sub-Gaussian with variance proxy $L^2$, as is every portfolio return $\mathbf{w}^\top r$ with $\|\mathbf{w}\|_2\le 1$, the quantity whose lower quantiles the CVaR program of Section~\ref{sec:phase5} optimizes. An analogous bound with exponential decay holds for log-concave $\mu$ \cite{ledoux2001concentration}, so exchanging a Gaussian prior for a uniform one does not evade it, and the argument extends to diffusion samplers, which are Lipschitz in the noise injected across the reverse chain. Jaini et al.~\cite{jaini2020tails} establish the corresponding statement for flow architectures and propose learning a heavier-tailed source in response.

No finite $L$ reproduces a target whose tails are heavier than Gaussian, and the calibrated Merton model is such a target despite being a mixture of Gaussians: its moment generating function $\exp(\tfrac{1}{2}\sigma^{2}\theta^{2}+\lambda(e^{\mu_J\theta+\sigma_J^{2}\theta^{2}/2}-1))$ grows doubly exponentially in $\theta$ and exceeds $e^{\theta^{2}s^{2}/2}$ for every finite $s$. This is an asymptotic statement and carries little force at the quantiles at which risk is measured, where a moderate $L$ suffices; it establishes that no exact match exists, and motivates asking not whether a target is reachable but how much distortion approximating it demands.

What varies is severity. Caffarelli's contraction theorem \cite{caffarelli2000monotonicity} gives $\mathrm{Lip}(T)\le\beta^{-1/2}$ for the Brenier map from a standard Gaussian to a target $e^{-W}$ with $D^2W\succeq\beta I$; heavy-tailed targets satisfy no such uniform log-concavity, and the map correspondingly expands. The Monge--Amp\`ere equation locates where. Since $DT=D^2u\succeq 0$ and $\det A\le\|A\|_{\mathrm{op}}^d$,
\begin{equation}
    \|DT(x)\|_{\mathrm{op}}\;\ge\;\big(f(x)/g(T(x))\big)^{1/d},
    \label{eq:lipschitz}
\end{equation}
so local distortion is bounded below by the density ratio and grows wherever the target density has a trough. Section~\ref{sec:phase1} evaluates this in closed form for the calibrated model.

SDOT is exempt because its map is not Lipschitz by construction. The map sends each cell $W_i(h^*)$ to $y_i$ and is discontinuous across cell boundaries, so tail observations are reached by a jump, and their frequency is inherited through \eqref{eq:optimality} rather than learned against a regularity constraint.

\section{Methodology}
\label{sec:method}
We apply the geometric variational framework of Section~\ref{sec:geometric} to financial scenario generation, with algorithmic enhancements from An et al.~\cite{an2020aeot}.

\paragraph{Algorithm 1 (MC Semi-Discrete OT Solver).}
In high dimensions ($d \gg 2$), exact computation of cell volumes is intractable because the spatial complexity of the convex hull of $N$ points in $\mathbb{R}^d$ is $O(N^{\lfloor d/2 \rfloor})$. Following An et al.~\cite{an2020aeot}, we estimate cell volumes via Monte Carlo sampling. At each iteration, $M$ Sobol quasi-Monte Carlo samples from $[-\tfrac{1}{2}, \tfrac{1}{2}]^d$ are assigned to their nearest cell via the assignment $i^*(x) = \arg\max_i \{ \langle x, y_i \rangle + h_i \}$, which reduces to a GPU-parallelized matrix multiply. The empirical cell volumes $\hat{w}_i = \#\{j \mid i^*(x_j) = i\} / M$ approximate the gradient~\eqref{eq:gradient}, and $h$ is updated via Adam with centering ($h \leftarrow h - \bar{h}$) to remove the gauge freedom. When convergence stalls as measured by $\|\nabla_h E\|_2$, the Monte Carlo batch size is doubled adaptively up to a configurable maximum. The MC estimation error decreases as $O(1/\sqrt{M})$, and the convergence of the energy is guaranteed by the strict convexity of~\eqref{eq:energy} on $\mathcal{H}$.

\paragraph{Algorithm 2 (Piecewise-Linear Generation).}
Following the extended OT map of An et al.~\cite{an2020aeot}, given the solved weights $h^*$, new samples are generated in three steps. First, uniform samples are drawn and their top-$K$ nearest cells are identified via the power diagram assignment. Second, cell pairs are filtered by a dihedral angle threshold on the lifted paraboloid, retaining only pairs whose supporting planes meet at a shallow angle and rejecting those separated by a steep one, so that interpolation is confined to regions where the map varies slowly. Third, paired source points are interpolated as $P_{\text{gen}} = (1-w) P_i + w P_j$, where $w$ controls the dissimilarity. This produces novel samples that lie on the piecewise-linear extension of the transport map, generating points between training observations rather than merely resampling them. The extension remains discontinuous, interpolation is confined within accepted pairs and the filtered boundaries are the jumps across which the map is not Lipschitz. It is on this piecewise-linear extension, rather than the atomic map $W_i(h^*)\mapsto y_i$, that the samples of Section~\ref{sec:experiments} are generated.

\section{Experimental Results}
\label{sec:experiments}
We evaluate SDOT against five baseline neural generators across five phases, from synthetic calibration to a 21-year out-of-sample portfolio backtest. All neural generators use hyperparameter sweeps over hidden sizes $\{128,256,512\}$ and learning rates. Phase~2 is the primary experiment: it is the only setting in which the distortion bound \eqref{eq:lipschitz} is available in closed form rather than estimated.

The neural baselines are: FM, flow matching \cite{lipman2023flow} with conditional optimal-transport paths; CNF, a free-form neural ODE trained by maximum likelihood \cite{chen2018neural,grathwohl2019ffjord}; OT-Flow \cite{onken2021otflow}; DDPM \cite{ho2020denoising} with $T=1{,}000$ denoising steps; and WGAN, a Wasserstein GAN with gradient penalty \cite{gulrajani2017wgangp}. All use a three-hidden-layer MLP backbone. Training is by Adam at batch size 512 for at most 2,000 epochs with early stopping (patience 30, evaluated every 10), and all draw from a standard Gaussian base.

Results report mean $\pm$ standard deviation across 5 random seeds unless otherwise noted. Experiments are run on a single V100 NVIDIA GPU with JAX 0.5.0 \cite{jax2018github}. We also include two non-neural baselines. Bootstrap (uniform resampling with replacement from the training data) reproduces the training measure exactly and therefore bounds the attainable fidelity to that measure. Gaussian (multivariate normal fit via sample mean and covariance) provides a parametric reference. Section~\ref{sec:phase5} additionally reports minimum-variance optimization, which minimizes portfolio variance under the same constraint set without using generated scenarios.

We use S\&P~500 daily OHLCV data spanning 1962--2026, comprising 3.6 million rows across 644 symbols, together with membership records that identify the exact periods each stock was an index constituent. Returns outside a stock's membership window are set to zero, preventing survivorship bias. A hard filter removes rows where $|r_t| > 0.95$, eliminating 669 delisted-ticker artifacts while preserving all legitimate crash events. This is discussed in more detail in Section~\ref{sec:ablation}. 

For controlled experiments (Phases~1--3, Sections~\ref{sec:phase1}--\ref{sec:phase3}), we calibrate a Merton jump-diffusion model~\cite{merton1976} to the S\&P~500 empirical distribution by moment-matching ($\sigma = 0.001$, $\lambda = 8.06$ annualized, $\Delta = 1/252$, $\mu_J = -0.006$, $\sigma_J = 0.132$, kurtosis 93.8). We report tail ratios $\text{tail}_{k\sigma} = \Prob_{\text{gen}}(|x| > k\hat{\sigma}) / \Prob_{\text{real}}(|x| > k\hat{\sigma})$ where $\hat{\sigma}$ is the standard deviation of the reference distribution, taken as the per-regime Merton value in Phase~2 and the empirical value elsewhere (ideal $= 1.0$). Distributional distances include energy distance, maximum mean discrepancy (MMD), and sliced Wasserstein-1 (SW1). For portfolio risk evaluation, we compute CVaR relative error $\epsilon_\alpha = |{\CVaR_\alpha^{\text{gen}} - \CVaR_\alpha^{\text{real}}}| / |{\CVaR_\alpha^{\text{real}}}|$ at the 1\% and 5\% levels across 100 random long-only portfolios.

For the backtest of Section~\ref{sec:phase5}, we employ the Rockafellar--Uryasev LP formulation for mean-CVaR optimization \cite{rockafellar2000optimization,rockafellar2002conditional}. The long-short setting decomposes weights as $\mathbf{w} = \mathbf{w}^+ - \mathbf{w}^-$ with constraints on net exposure ($=0$), gross leverage ($\leq L_{\text{gross}}$), and per-stock weight ($\leq w_{\max}$). The LP is solved via HiGHS through \texttt{scipy} with sparse constraint matrices \cite{huangfu2018parallelizing}. With $S=10{,}000$ scenarios and $D=100$ assets, the problem has $10{,}201$ variables and solves in under one second.

\subsection{Phase 1 -- Synthetic Calibration}
\label{sec:phase1}
Moment-matching via Nelder--Mead on closed-form moment formulas achieves near-exact fit (loss $= 2.53\times10^{-10}$), with kurtosis 93.8 against the empirical 87.7.

Because the target factorizes across coordinates, the Brenier map from $\mathcal{N}(0,I_d)$ decomposes into $d$ copies of the monotone rearrangement $T_1 = G^{-1}\circ\Phi$, and \eqref{eq:lipschitz} holds with equality: $T_1'(x) = \varphi(x)/g(T_1(x))$. The global supremum diverges in all four regimes of Table~\ref{tab:regimes}, consistent with Section~\ref{sec:lipschitz}, since the Merton density is a Poisson mixture whose $k$-th component has variance $\sigma^2 + k\sigma_J^2$ and therefore has tails heavier than any Gaussian. We report instead the distortion at the interior density minimum, attained at $r^*$ between $-4\times10^{-4}$ and $-5\times10^{-4}$ in every regime, six to eight core standard deviations ($\sigma\sqrt{\Delta} = 6.3\times10^{-5}$) from the origin and hence at the boundary between the diffusive and jump regimes rather than in the far tail. Its sign is inherited from $\mu_J<0$: the map's greatest local distortion falls on the loss side of the distribution, which is the side the CVaR objective of Section~\ref{sec:phase5} depends on.

\subsection{Phase 2 -- Sweeping the Required Distortion}
\label{sec:phase2}
We evaluate all generators on 10,000 IID samples in $\mathbb{R}^{100}$ drawn from the calibrated Merton model across four regimes (Table~\ref{tab:regimes}), constructed by reducing jump frequency and increasing jump magnitude to span kurtosis from 94 to 1,679. The target has full support and a convex domain, so the topological obstruction of Section~\ref{sec:regularity} is absent by construction and any failure observed is attributable to the metric one. What the sweep varies is the distortion the Brenier map must supply. Across the four regimes the density trough deepens sixty-eight-fold and the local distortion at that trough rises six-fold, while the ground-truth distribution remains known exactly.

\begin{table}[t]
\centering
\caption{Merton jump-diffusion regimes for the Phase~2.
$g_{\min}$ is the density at the trough separating the diffusive core from the
jump component, and $L^*$ is the local distortion of the Brenier map from
$\mathcal{N}(0,1)$ at that trough, normalized by the target standard deviation.
The global Lipschitz constant is infinite in all four regimes, these columns report the severity of the obstruction over the range at which tail risk is measured.}
\label{tab:regimes}
\small
\begin{tabular}{@{}lcccrrr@{}}
\toprule
Regime & $\lambda$ & $\mu_J$ & $\sigma_J$ & Kurtosis & $g_{\min}$ & $L^*/\hat{\sigma}$ \\
\midrule
Base (calibrated) & 8.06 & $-0.006$ & 0.132 & 93.8    & $9.5{\times}10^{-2}$ & 18.2 \\
Moderate          & 2.0  & $-0.05$  & 0.3   & 377.8   & $1.0{\times}10^{-2}$ & 46.6 \\
Heavy             & 1.0  & $-0.05$  & 0.4   & 755.9   & $3.9{\times}10^{-3}$ & 68.8 \\
Extreme           & 0.45 & $-0.08$  & 0.5   & 1{,}679 & $1.4{\times}10^{-3}$ & 112.5 \\
\bottomrule
\end{tabular}
\end{table}
Table~\ref{tab:phase2} presents the key metrics across all four regimes, and both quantities move as Section~\ref{sec:lipschitz} predicts. SDOT's tail ratios remain in $[0.85, 0.94]$ with standard deviations below 0.025, as expected of a map to which no distortion bound applies. Every learned generator falls on one of two branches. Both branches are consistent with \eqref{eq:subgaussian}: a generator whose Lipschitz constant remains small stays sub-Gaussian with a small variance proxy and compresses, while one driven to a large constant is sub-Gaussian with a proxy so large that at $3\sigma$ and $4\sigma$ it over-produces, and pays for it in cross-seed variance. Those whose Lipschitz constant falls short of the requirement compress: OT-Flow and Gaussian reach tail$_{4\sigma}$ of only $0.10$ and $0.29$ at the extreme regime. Those that overshoot inflate, and do so with a cross-seed standard deviation that grows alongside the mean.

CNF inflates from $0.96\pm1.06$ to $311\pm34.0$ and DDPM from $1.32\pm0.09$ to $48.7\pm10.0$, with WGAN and FM following the same pattern at smaller magnitude. SDOT's energy distance declines as tails get heavier (0.000916 at base to 0.000109 at extreme), as does Bootstrap's, indicating a scale effect of the metric under increasing dispersion rather than a property of either method. At the extreme regime, SDOT's energy distance ($1.09 \times 10^{-4}$) is $169\times$ better than FM ($1.84 \times 10^{-2}$) and $5{,}780\times$ better than CNF ($6.30 \times 10^{-1}$). 

At base and moderate severity, where the required distortion is smallest, FM attains tail ratios closer to $1.0$ than SDOT ($1.12$ and $1.01$ against $0.88$ and $0.94$). This is the expected outcome. At low severity a Lipschitz map approximates the target well, and SDOT's residual bias comes from the piecewise-linear extension of Section~\ref{sec:method} rather than from any constraint on the map. The separation appears as severity grows, where FM reaches $4.88 \pm 1.94$ and SDOT holds at $0.91 \pm 0.02$.

\begin{table}[t]
\centering
\caption{Phase 2 -- Merton regularity sweep (mean $\pm$ std across 5 seeds). SDOT maintains stable tail ratios across all regimes while every learned generator diverges in tail ratio, cross-seed variance, or both. Energy, MMD, and SW1 are scaled by $10^3$.}
\label{tab:phase2}
\small
\setlength{\tabcolsep}{3pt}
\begin{tabular}{@{}l rr rrr@{}}
\toprule
Method & tail$_{3\sigma}$ & tail$_{4\sigma}$ & Energy & MMD & SW1 \\
\midrule
\multicolumn{6}{@{}l}{\emph{Base} (kurtosis 93.8)} \\
SDOT (ours) & $0.88{\pm}0.01$ & $0.85{\pm}0.01$ & $0.92{\pm}0.20$ & $1.77{\pm}0.50$ & $1.75{\pm}0.07$ \\
Bootstrap   & $1.00{\pm}0.01$ & $1.00{\pm}0.01$ & $0.28{\pm}0.02$ & $-0.05{\pm}0.04$ & $0.34{\pm}0.01$ \\
Gaussian    & $0.16{\pm}0.00$ & $0.01{\pm}0.00$ & $4.37{\pm}0.36$ & $0.52{\pm}0.14$ & $3.71{\pm}0.04$ \\
FM          & $1.12{\pm}0.03$ & $1.08{\pm}0.03$ & $2.80{\pm}0.51$ & $2.28{\pm}0.45$ & $2.42{\pm}0.17$ \\
CNF         & $0.96{\pm}1.06$ & $0.70{\pm}1.00$ & $25.0{\pm}36.8$ & $49.4{\pm}64.4$ & $7.57{\pm}5.27$ \\
OT-Flow     & $0.64{\pm}0.21$ & $0.28{\pm}0.18$ & $21.9{\pm}21.0$ & $34.0{\pm}35.4$ & $6.34{\pm}2.45$ \\
DDPM        & $1.32{\pm}0.09$ & $1.04{\pm}0.07$ & $19.2{\pm}2.3$ & $25.9{\pm}3.7$ & $14.6{\pm}0.5$ \\
WGAN        & $1.19{\pm}1.28$ & $0.44{\pm}0.58$ & $8.24{\pm}13.1$ & $10.6{\pm}16.5$ & $4.27{\pm}4.37$ \\
\midrule
\multicolumn{6}{@{}l}{\emph{Moderate} (kurtosis 377.8)} \\
SDOT (ours) & $0.94{\pm}0.01$ & $0.93{\pm}0.00$ & $0.37{\pm}0.06$ & $0.15{\pm}0.18$ & $1.04{\pm}0.11$ \\
Bootstrap   & $1.00{\pm}0.01$ & $1.00{\pm}0.01$ & $0.26{\pm}0.06$ & $-0.07{\pm}0.09$ & $0.41{\pm}0.02$ \\
Gaussian    & $0.56{\pm}0.03$ & $0.02{\pm}0.00$ & $42.2{\pm}1.3$ & $8.83{\pm}0.98$ & $11.5{\pm}0.1$ \\
FM          & $1.01{\pm}0.08$ & $1.02{\pm}0.07$ & $10.4{\pm}1.5$ & $4.27{\pm}1.67$ & $4.37{\pm}0.61$ \\
CNF         & $81.7{\pm}5.9$ & $65.4{\pm}7.3$ & $762{\pm}74.6$ & $337{\pm}14.1$ & $71.2{\pm}8.4$ \\
OT-Flow     & $0.34{\pm}0.04$ & $0.02{\pm}0.00$ & $41.0{\pm}1.3$ & $13.7{\pm}1.1$ & $10.9{\pm}0.2$ \\
DDPM        & $5.56{\pm}1.81$ & $3.74{\pm}1.23$ & $36.5{\pm}13.4$ & $26.4{\pm}13.2$ & $17.3{\pm}5.1$ \\
WGAN        & $3.82{\pm}1.63$ & $1.40{\pm}0.51$ & $25.1{\pm}19.4$ & $11.8{\pm}11.2$ & $8.01{\pm}3.93$ \\
\midrule
\multicolumn{6}{@{}l}{\emph{Heavy} (kurtosis 755.9)} \\
SDOT (ours) & $0.93{\pm}0.02$ & $0.93{\pm}0.02$ & $0.28{\pm}0.14$ & $0.22{\pm}0.64$ & $0.75{\pm}0.17$ \\
Bootstrap   & $1.01{\pm}0.01$ & $1.01{\pm}0.01$ & $0.18{\pm}0.07$ & $-0.01{\pm}0.29$ & $0.39{\pm}0.02$ \\
Gaussian    & $1.21{\pm}0.02$ & $0.07{\pm}0.01$ & $68.2{\pm}2.5$ & $32.7{\pm}3.9$ & $14.4{\pm}0.2$ \\
FM          & $1.40{\pm}0.11$ & $1.24{\pm}0.09$ & $14.8{\pm}1.4$ & $9.32{\pm}1.47$ & $5.04{\pm}0.27$ \\
CNF         & $160{\pm}32.1$ & $109{\pm}39.5$ & $726{\pm}149$ & $343{\pm}18.3$ & $71.3{\pm}13.9$ \\
OT-Flow     & $0.60{\pm}0.06$ & $0.03{\pm}0.00$ & $64.7{\pm}1.4$ & $35.3{\pm}2.9$ & $13.4{\pm}0.1$ \\
DDPM        & $9.07{\pm}5.24$ & $5.61{\pm}3.54$ & $35.9{\pm}10.5$ & $32.9{\pm}8.9$ & $9.76{\pm}3.76$ \\
WGAN        & $3.04{\pm}3.15$ & $1.25{\pm}1.58$ & $25.6{\pm}5.5$ & $16.2{\pm}5.6$ & $8.12{\pm}1.15$ \\
\midrule
\multicolumn{6}{@{}l}{\emph{Extreme} (kurtosis 1,679)} \\
SDOT (ours) & $0.91{\pm}0.02$ & $0.89{\pm}0.02$ & $0.11{\pm}0.01$ & $0.08{\pm}0.08$ & $0.75{\pm}0.12$ \\
Bootstrap   & $1.00{\pm}0.02$ & $1.00{\pm}0.02$ & $0.10{\pm}0.01$ & $-0.10{\pm}0.02$ & $0.31{\pm}0.01$ \\
Gaussian    & $3.40{\pm}0.35$ & $0.29{\pm}0.06$ & $83.5{\pm}0.8$ & $75.5{\pm}1.6$ & $14.4{\pm}0.2$ \\
FM          & $4.88{\pm}1.94$ & $2.76{\pm}0.80$ & $18.4{\pm}3.1$ & $98.5{\pm}4.9$ & $5.01{\pm}0.69$ \\
CNF         & $311{\pm}34.0$ & $201{\pm}49.1$ & $630{\pm}67.4$ & $357{\pm}33.9$ & $58.5{\pm}6.8$ \\
OT-Flow     & $1.56{\pm}0.30$ & $0.10{\pm}0.04$ & $78.2{\pm}0.9$ & $82.7{\pm}2.1$ & $13.1{\pm}0.2$ \\
DDPM        & $48.7{\pm}10.0$ & $35.0{\pm}7.5$ & $49.4{\pm}11.1$ & $109{\pm}6.8$ & $16.1{\pm}3.1$ \\
WGAN        & $10.9{\pm}2.41$ & $6.80{\pm}1.98$ & $14.2{\pm}2.4$ & $341{\pm}58.1$ & $5.88{\pm}0.80$ \\
\bottomrule
\end{tabular}
\end{table}

\subsection{Phase 3 -- Cross-Domain Transfer}
\label{sec:phase3}
We evaluate whether generators trained on synthetic Merton data (Phase~2 base regime, 10,000 samples $\times$ 100D) transfer to real S\&P~500 daily returns from 1962--2026. For each of 5 seeds, 100 symbols are randomly selected from active S\&P~500 constituents. This tests the practical scenario where a risk model calibrated to a parametric distribution must produce scenarios for actual market data.

Table~\ref{tab:phase34} shows the combined results. Methods separate into two tiers. The tail-preserving group (SDOT, Bootstrap, FM, DDPM) produces meaningful tail mass at 4$\sigma$ and beyond. The tail-deficient group (WGAN, OT-Flow, Gaussian, CNF) has tails that decay to near-zero by 5$\sigma$, with GAN and OT-Flow both reaching $0.03$ at 6$\sigma$. SDOT achieves the best tail$_{3\sigma}$ (1.17, closest to 1.0) and the best energy distance and MMD among tail-preserving methods. CNF achieves the best overall energy distance (0.0136) but has the worst CVaR error among stable methods (62.9\% at the 1\% level), demonstrating the disconnect between bulk distributional accuracy and tail risk estimation.

\begin{table*}[t]
\centering
\caption{Phases 3 and 4 -- cross-domain transfer and real held-out
evaluation, mean $\pm$ std across 5 seeds. Tail ratios ideal $= 1.0$;
energy, MMD, and SW1 scaled by $10^3$; CVaR relative errors
$\epsilon_\alpha$ are means, in percent; all lower better except tail ratios.}
\label{tab:phase34}
\small
\begin{tabular}{@{}l rrrr rrr rr@{}}
\toprule
& \multicolumn{4}{c}{Tail ratios} & \multicolumn{3}{c}{Distances} & \multicolumn{2}{c}{CVaR error} \\
\cmidrule(lr){2-5} \cmidrule(lr){6-8} \cmidrule(lr){9-10}
Method & $3\sigma$ & $4\sigma$ & $5\sigma$ & $6\sigma$ & Energy & MMD & SW1 & $\epsilon_{1\%}$ & $\epsilon_{5\%}$ \\
\midrule
\multicolumn{10}{@{}l}{\underline{\emph{Phase 3}}} \\
SDOT (ours) & $1.17{\pm}0.11$ & $1.95{\pm}0.15$ & $2.71{\pm}0.33$ & $3.40{\pm}0.70$ & $14.9{\pm}1.4$ & $46.3{\pm}5.2$ & $6.85{\pm}0.31$ & $21.8{\pm}1.9$ & $13.6{\pm}1.1$ \\
Bootstrap & $1.33{\pm}0.13$ & $2.28{\pm}0.18$ & $3.27{\pm}0.38$ & $4.23{\pm}0.82$ & $20.1{\pm}1.0$ & $58.9{\pm}4.7$ & $8.50{\pm}0.31$ & $17.0{\pm}1.7$ & $12.8{\pm}1.8$ \\
Gaussian & $0.24{\pm}0.12$ & $0.02{\pm}0.02$ & $0.00{\pm}0.00$ & $0.00{\pm}0.00$ & $41.1{\pm}1.7$ & $64.8{\pm}4.2$ & $11.0{\pm}0.3$ & $42.7{\pm}2.2$ & $18.9{\pm}1.8$ \\
FM & $1.49{\pm}0.14$ & $2.50{\pm}0.20$ & $3.46{\pm}0.44$ & $4.29{\pm}0.91$ & $31.7{\pm}1.6$ & $64.1{\pm}4.9$ & $9.96{\pm}0.31$ & $20.6{\pm}1.5$ & $13.7{\pm}1.6$ \\
CNF & $0.11{\pm}0.06$ & $0.01{\pm}0.01$ & $0.00{\pm}0.00$ & $0.00{\pm}0.00$ & $13.6{\pm}0.8$ & $24.5{\pm}3.0$ & $5.19{\pm}0.20$ & $62.9{\pm}1.7$ & $47.0{\pm}2.3$ \\
OT-Flow & $0.62{\pm}0.16$ & $0.28{\pm}0.11$ & $0.09{\pm}0.05$ & $0.03{\pm}0.02$ & $43.3{\pm}1.5$ & $72.8{\pm}4.5$ & $10.5{\pm}0.2$ & $33.5{\pm}2.4$ & $14.4{\pm}1.7$ \\
DDPM & $1.55{\pm}0.09$ & $2.14{\pm}0.19$ & $2.78{\pm}0.36$ & $3.55{\pm}0.62$ & $55.0{\pm}2.7$ & $122.9{\pm}8.1$ & $21.4{\pm}0.3$ & $243.7{\pm}10.9$ & $167.8{\pm}7.2$ \\
WGAN & $0.63{\pm}0.15$ & $0.28{\pm}0.11$ & $0.11{\pm}0.06$ & $0.03{\pm}0.02$ & $18.8{\pm}1.3$ & $53.0{\pm}5.4$ & $7.36{\pm}0.32$ & $31.5{\pm}2.7$ & $16.6{\pm}1.4$ \\
\midrule
\multicolumn{10}{@{}l}{\underline{\emph{Phase 4}}} \\
SDOT (ours) & $0.33{\pm}0.03$ & $0.28{\pm}0.04$ & $0.27{\pm}0.07$ & $0.27{\pm}0.09$ & $8.89{\pm}1.14$ & $21.4{\pm}3.5$ & $4.94{\pm}0.51$ & $51.3{\pm}3.3$ & $49.4{\pm}3.0$ \\
Bootstrap & $0.38{\pm}0.03$ & $0.34{\pm}0.03$ & $0.33{\pm}0.05$ & $0.32{\pm}0.06$ & $9.11{\pm}1.38$ & $23.0{\pm}3.8$ & $5.00{\pm}0.51$ & $51.2{\pm}3.0$ & $51.2{\pm}2.8$ \\
Gaussian & $0.21{\pm}0.16$ & $0.20{\pm}0.22$ & $0.21{\pm}0.25$ & $0.21{\pm}0.26$ & $3.34{\pm}0.52$ & $11.0{\pm}2.3$ & $3.45{\pm}0.29$ & $70.8{\pm}1.6$ & $59.3{\pm}2.2$ \\
FM & $0.30{\pm}0.08$ & $0.22{\pm}0.08$ & $0.18{\pm}0.08$ & $0.16{\pm}0.09$ & $6.75{\pm}2.04$ & $19.8{\pm}7.1$ & $4.44{\pm}0.79$ & $57.9{\pm}6.9$ & $54.5{\pm}6.7$ \\
CNF & $7.9{\pm}7.5$ & $11.6{\pm}12.7$ & $14.6{\pm}16.6$ & $18.4{\pm}22.7$ & $260{\pm}339$ & $185{\pm}121$ & $32.8{\pm}30.1$ & $46.4{\pm}18.8$ & $61.3{\pm}24.6$ \\
OT-Flow & $0.21{\pm}0.17$ & $0.19{\pm}0.22$ & $0.20{\pm}0.26$ & $0.20{\pm}0.27$ & $5.26{\pm}1.08$ & $17.3{\pm}5.8$ & $4.17{\pm}0.48$ & $82.0{\pm}2.2$ & $74.9{\pm}3.1$ \\
DDPM & $1.23{\pm}0.60$ & $1.42{\pm}0.83$ & $1.61{\pm}1.08$ & $1.78{\pm}1.30$ & $9.10{\pm}2.70$ & $18.8{\pm}5.3$ & $5.44{\pm}1.52$ & $37.5{\pm}20.5$ & $42.8{\pm}19.1$ \\
WGAN & $0.66{\pm}0.25$ & $0.60{\pm}0.27$ & $0.53{\pm}0.28$ & $0.46{\pm}0.27$ & $9.18{\pm}2.77$ & $16.0{\pm}4.1$ & $4.78{\pm}0.86$ & $55.1{\pm}20.2$ & $50.6{\pm}19.0$ \\
\bottomrule
\end{tabular}
\end{table*}

\subsection{Phase 4 -- Real S\&P 500 Held-Out Evaluation}
\label{sec:phase4}

We train all generators on real S\&P~500 daily returns from 1962--2004 and evaluate on a held-out test period of 2005--2026. For each of 5 seeds, 100 symbols are randomly selected from constituents active at the train/test boundary. This experiment isolates transfer under regime change rather than testing the mechanism of Section~\ref{sec:lipschitz}, since the training and evaluation measures differ. The test period includes the 2008 financial crisis, the 2020 COVID crash, and the post-quantitative-easing environment, all of which differ fundamentally from the training data.

Table~\ref{tab:phase34} presents the complete results. All methods exhibit CVaR errors of 38--82\%, reflecting the difficulty of predicting post-crisis tail dynamics from pre-crisis training data. Bootstrap reproduces the training distribution exactly and still attains a tail$_{3\sigma}$ of only $0.38$, so the low tail ratios of SDOT, Bootstrap, and most other models reflect nonstationarity rather than a failure to represent the training data. Section~\ref{sec:phase5} addresses this through annual expanding-window retraining.

Bootstrap and SDOT are statistically indistinguishable in this experiment, which is the expected outcome given that both reproduce the training measure faithfully. They differ in the form that measure takes. The optimality condition $w_i(h^*) = 1/N$ gives every observation a power cell of equal source measure, so SDOT spreads each $1/N$ continuously over a region of positive volume where Bootstrap concentrates it at the observed point. Section~\ref{sec:phase5} shows this difference matters downstream, the CVaR program finds no non-trivial portfolio on an atomic scenario set.

\subsubsection{Data Cleaning Ablation}
\label{sec:ablation}
Preprocessing shapes the tails these models are meant to capture. Winsorization collapses a $-45\%$ crash and a $-99\%$ artifact to the same boundary value, destroying the ordering a tail-sensitive model must learn, while hard filtering preserves the magnitude of everything below the threshold. Table~\ref{tab:ablation} reports kurtosis under each setting on the Phase~4 test set (2005--2026, 100 randomly selected constituents, mean $\pm$ std across 5 seeds). We show that winsorization is unsuitable at every threshold, at the 99.9th percentile kurtosis falls to 10.6, (near-Gaussian) erasing the tail structure that differentiates generators. The 99.99th percentile and above yields values 900 and higher, because clipping to boundary values creates artificial density spikes. 

Among hard filters, $|r| > 1.0$ is too permissive (kurtosis $262.6 \pm 135.8$, cross-seed std exceeding half the mean), retaining delisted-ticker artifacts such as $-99.97\%$ returns from residual price records; $|r| > 0.50$ and $0.75$ are too aggressive (33.9 and 54.7), leaving so little tail mass that all generators converge and the experiment loses discriminative power. We filter at $|r| > 0.95$ which balances considerations above, and this threshold is used throughout our experiments. It produces kurtosis of 87.7 $\pm$ 25.5. All real crash events (2008 GFC, 2020 COVID) are preserved, and enough tail structure remains for the generative models to differentiate themselves.

\begin{table}[t]
\centering
\caption{Data cleaning ablation on the Phase 4 test period (2005--2026). Kurtosis under different preprocessing thresholds.}
\label{tab:ablation}
\small
\begin{tabular}{@{}lrr@{}}
\toprule
Method & Rows affected & Kurtosis \\
\midrule
Baseline (raw) & 0 & $69{,}245 \pm 41{,}552$ \\
Winsorize 99.9th & 7,166 clipped & $10.6 \pm 0.4$ \\
Winsorize 99.99th & 720 clipped & $900.2 \pm 420.0$ \\
Hard $|r| > 1.00$ & 409 dropped & $262.6 \pm 135.8$ \\
Hard $|r| > 0.95$ (used) & 669 dropped & $87.7 \pm 25.5$ \\
Hard $|r| > 0.75$ & 768 dropped & $54.7 \pm 10.4$ \\
Hard $|r| > 0.50$ & 932 dropped & $33.9 \pm 4.8$ \\
\bottomrule
\end{tabular}
\end{table}

\subsection{Phase 5 -- Portfolio Backtest}
\label{sec:phase5}
We conduct an out-of-sample portfolio backtest over 2005--2026 (258 months, 5,406 trading days) with monthly rebalancing and annual generator retraining. Each year, all generators are retrained on the expanding window of all available historical data up to that point. A model-agnostic screen selects 100 stocks for training and evaluation of each model. A multivariate normal is fit to all active constituents of the S\&P~500 universe, 10,000 scenarios are generated, mean-CVaR optimization is run on the full universe, and the top 100 stocks by absolute weight magnitude are selected. 

The long-short constraint set enforces zero net exposure, $2\times$ gross leverage, and a 5\% maximum weight per stock, with transaction costs of 5 basis points proportional to turnover. Under CVaR-dominant optimization with an inequality leverage constraint the zero portfolio is feasible, so the program takes a position only when a generator's scenarios contain structure worth the tail risk. Zero net exposure hedges market beta, leaving performance dependent on the tail risk estimates for individual long-short pairs. These are lower quantiles of $\mathbf{w}^\top r$, a Lipschitz functional under the leverage constraint and hence the object bounded by \eqref{eq:subgaussian}: a generator with sub-Gaussian scenarios understates the loss quantile the program minimizes, making positions appear safer than they are.

Table~\ref{tab:phase5} separates the two properties a scenario set requires. Bootstrap and Gaussian yield $\mathbf{w}=\mathbf{0}$ at every rebalance and are excluded, for opposite reasons: Bootstrap is tail-faithful but atomic, so no long-short pair beats the zero portfolio, while Gaussian is absolutely continuous but sub-Gaussian by construction and offers no tail structure to act on. SDOT has both properties, spreading each $1/N$ over a cell of positive volume while inheriting tail frequencies through \eqref{eq:optimality}, and produces the best risk-adjusted market-neutral strategy: Sharpe 0.70, Sortino 1.16, and maximum drawdown $-2.60\%$, against $-35.7\%$ for OT-Flow, $-33.0\%$ for minimum variance, and $-55.2\%$ for SPY. CNF and WGAN are the only other generators with positive risk-adjusted returns (Sharpe 0.40 and 0.21), both with substantially larger drawdowns.

\begin{table}[t]
\centering
\caption{Phase 5 -- Long-short portfolio backtest (2005--2026). Zero net exposure, $2\times$ gross leverage, 5\% max weight, 5bps transaction costs. Bootstrap and Gaussian yield $\mathbf{w} = \mathbf{0}$ at every rebalance. SPY shown as market reference and excluded from ranking. All figures except Sharpe and Sortino in percent. Higher is better for Sharpe, Sortino, and CAGR; for Vol, Max DD, and CVaR$_{5\%}$, closer to zero is better. Best result in \textbf{bold}, second best \underline{underlined}.}
\label{tab:phase5}
\small
\begin{tabular}{@{}l rr rrrr@{}}
\toprule
Strategy & Sharpe & Sortino & CAGR & Vol & Max DD & CVaR$_{5\%}$ \\
\midrule
SDOT (ours)    & $\mathbf{0.70}$ & $\mathbf{1.16}$ & \underline{$0.95$} & $1.36$ & $\mathbf{-2.60}$ & $-0.18$ \\
CNF            & \underline{$0.40$} & \underline{$0.62$} & $0.41$ & \underline{$1.06$} & \underline{$-7.52$} & \underline{$-0.15$} \\
WGAN           & $0.21$ & $0.32$ & $\mathbf{1.45}$ & $9.63$ & $-28.4$ & $-1.30$ \\
Min.\ Var. & $0.04$ & $0.06$ & $0.01$ & $6.82$ & $-33.0$ & $-0.96$ \\
DDPM           & $-0.05$ & $-0.07$ & $-0.13$ & $1.98$ & $-14.5$ & $-0.34$ \\
OT-Flow        & $-0.12$ & $-0.18$ & $-1.18$ & $6.86$ & $-35.7$ & $-1.01$ \\
FM             & $-0.23$ & $-0.30$ & $-0.23$ & $\mathbf{0.95}$ & $-11.1$ & $\mathbf{-0.14}$ \\
\midrule
SPY & $0.64$ & $0.90$ & $10.87$ & $18.95$ & $-55.2$ & $-2.90$ \\
\bottomrule
\end{tabular}
\end{table}

\section{Limitations and Future Work}
\label{sec:limitations}
Two limitations bound the scope of the comparison. First, every generator benchmarked here draws from a Gaussian base, learning a heavier-tailed source \cite{jaini2020tails,huster2021pareto} is the complementary remedy to relaxing the map, and the two address different terms of the same obstruction rather than competing. 

Second, extending the power diagram to conditional or path-valued targets would supply the temporal structure these comparisons require, and the adversarial setting is the most promising route. An et al.~\cite{an2020aeotgan} explore this approach. Applied to scenario generation, this would pair SDOT's inherited tail frequencies with an adversarial critic that sharpens structure within cells. Additionally, Tail-GAN~\cite{cont2023tailgan} defines its scoring rule on the VaR and expected shortfall of a predefined set of benchmark strategies, so extending our Wasserstein GAN~\cite{gulrajani2017wgangp} with this scoring rule may lead to stronger results.

\section{Conclusion}
\label{sec:conclusion}

We have shown that generative models which learn transport from a simple base distribution, including normalizing flows, continuous normalizing flows, OT-flows, denoising diffusion, and adversarial networks, systematically fail to capture the heavy tails of financial return distributions. These models fail because they are Lipschitz, not because they are smooth or invertible: the Monge--Amp\`ere equation bounds a map's local distortion below by the density ratio $f/(g\circ T)$, which grows where the target density has a trough, and meeting that demand costs estimator stability. The constraint is architectural rather than a deficiency of training.

Semi-Discrete Optimal Transport relaxes the map's regularity rather than the source's tail class, reaching tail observations by crossing a cell boundary rather than by stretching. Across four regimes spanning kurtosis 94 to 1,679, over which the required distortion rises six-fold, SDOT holds tail ratios at $0.85$--$0.94$ with the lowest cross-seed variance of any method tested, and in a 21-year out-of-sample backtest yields the best risk-adjusted long-short strategy under CVaR-minimizing optimization. For risk applications the relevant property is not tractable likelihood or a smooth map, but a construction whose tail frequencies are inherited from the data rather than learned against a regularity constraint.

\bibliographystyle{ACM-Reference-Format}
\bibliography{bib} 


\end{document}